\documentclass[
  journal=pasa,
  manuscript=article-type,
  year=2020,
  volume=37,
]{cup-journal}
\usepackage{aas-macros} % PASA wants this
\makeatletter
\renewcommand\@biblabel[1]{[#1]}% plain brackets, no \textbf
\makeatother
\usepackage{rotating}
\usepackage{amsmath}
\usepackage{longtable}
\usepackage{amssymb}
\usepackage{hyperref}
\usepackage[nopatch]{microtype}
\usepackage{booktabs}
\usepackage{lscape}
\usepackage{subfig}
\usepackage{supertabular,booktabs}
\usepackage{gensymb}
\usepackage{float}
\usepackage{lscape}
\usepackage{aas-macros}
\usepackage{xcolor}

\usepackage{CJKutf8}

\newcommand{\source}{VAST~J172812.1$-$460801}
\newcommand{\fermi}{4FGL~J1728.0$-$4606}
\newcommand{\gaia}{\textit{Gaia}~DR3~5951944861092533248}
\newcommand{\gaiaone}{\textit{Gaia1}}
\newcommand{\psr}{PSR~J1728$-$4608}

\title{Discovery of the redback millisecond pulsar \psr\ with ASKAP}

\email{flora.petrou@postgrad.curtin.edu.au,yuanmingwang@swin.edu.au}

\author{F. Petrou}
\affiliation{International Centre for Radio Astronomy Research, Curtin University, Bentley, WA, Australia}
\author{Y. Wang}
\affiliation{Centre for Astrophysics and Supercomputing, Swinburne University of Technology, Hawthorn, VIC 3122, Australia}
\alsoaffiliation{ARC Centre of Excellence for Gravitational Wave Discovery (OzGrav), Hawthorn, VIC 3122, Australia}

\author{N. Hurley-Walker}
\affiliation{International Centre for Radio Astronomy Research, Curtin University, Bentley, WA, Australia}

\author{S. McSweeney}
\affiliation{International Centre for Radio Astronomy Research, Curtin University, Bentley, WA, Australia}

\author{L.Zhang}
\affiliation{State Key Laboratory of Radio Astronomy and Technology, National Astronomical Observatories, Chinese Academy of Sciences, Beijing 100101, China}
\alsoaffiliation{Centre for Astrophysics and Supercomputing, Swinburne University of Technology, VIC, 3122, Australia}

\author{R. G. Key}
\affiliation{Centre for Astrophysics and Supercomputing, Swinburne University of Technology, VIC, 3122, Australia}
\alsoaffiliation{University of Technology, Melbourne, Victoria, 3122, Australia}

\author{J. Freeburn}
\affiliation{Department of Physics and Astronomy, University of North Carolina at Chapel Hill, Chapel Hill, NC 27599-3255, USA}

\author{B. W. Meyers}
\affiliation{International Centre for Radio Astronomy Research, Curtin University, Bentley, WA, Australia}

\author{David L. Kaplan}
\affiliation{Department of Physics, University of Wisconsin-Milwaukee, P.O. Box 413, Milwaukee, WI 53201, USA}

\author{A. Zic}
\affiliation{Australia Telescope National Facility, CSIRO, Space and Astronomy, PO Box 76, Epping, NSW 1710, Australia}

\author{Tara Murphy}
\affiliation{Sydney Institute for Astronomy, School of Physics, The University of Sydney, New South Wales 2006, Australia}
\alsoaffiliation{ARC Centre of Excellence for Gravitational Wave Discovery (OzGrav), Hawthorn, VIC 3122, Australia}

\author{D. Dobie}
\affiliation{Sydney Institute for Astronomy, School of Physics, The University of Sydney, New South Wales 2006, Australia}
\alsoaffiliation{ARC Centre of Excellence for Gravitational Wave Discovery (OzGrav), Hawthorn, VIC 3122, Australia}

\author{Y. Maan}
\affiliation{National Centre for Radio Astrophysics, Tata Institute of Fundamental Research, Post Bag 3, Ganeshkhind, Pune - 411007, India}

\begin{document}

\begin{abstract}

We present the discovery of \psr, a new redback spider pulsar identified in images from the Australian SKA Pathfinder telescope. \psr\ is a millisecond pulsar with a spin period of 2.86\,ms, in a 5.05\,hr orbit with a companion star. The pulsar exhibits a radio spectrum of the form $S_\nu \propto \nu^\alpha$, with a measured spectral index of $\alpha = -1.8(3)$. It is eclipsed for 42\% of its orbit at 888\,MHz, and multi--frequency image--domain observations show that the egress duration scales with frequency as a power law with index $n = -1.74$, where longer duration eclipses are seen at lower frequencies. An optical counterpart is detected in archival Gaia data within $0.5''$ of the radio position. It has a mean G-band magnitude of 
18.8 mag and its light curve displays characteristics consistent with a combination of ellipsoidal modulation and irradiation effects. We also report the nearest \textit{Fermi} $\gamma$-ray source, located 2$'$ away from our source, as a possible association. A radio timing study constrains the intrinsic and orbital properties of the system, revealing orbital period variations that we attribute to changes in the gravitational quadrupole moment of the companion star. At the eclipse boundary, we measure a maximum dispersion measure excess of $2.0 \pm 1.2 \ \mathrm{pc\ cm^{-3}}$, corresponding to an electron column density of $5.9 \pm 3.6 \times10^{18} \ \mathrm{cm^{-2}}$. Modelling of the eclipse mechanism suggests that synchrotron absorption is the dominant cause of the eclipses observed at radio wavelengths. The discovery and characterisation of systems like \psr\ provide valuable insights into pulsar recycling, binary evolution, the nature of companion-driven eclipses, and the interplay between compact objects and their plasma environments.

\end{abstract}



\section{Introduction} \label{sec:intro}

Spider millisecond pulsars (MSPs) are a class of pulsars once considered rare, found in compact binary systems with orbital periods of less than a day and low-mass companion stars \citep{1988Natur.333..237F, 2013IAUS..291..127R, 2021ApJ...922L..13W}. They are further divided into two subclasses depending on the mass of their companions: redbacks (RBs) with semi-degenerate companions (0.3--0.7$M_\odot$) and black widows (BWs) with non-degenerate companions ($<\mathrm{0.1}M_\odot$) \citep{2013IAUS..291..127R}. As of August 2025, 32 confirmed RB pulsars and 49 BW pulsars have been identified in the Galactic field \citep[see SpiderCat;][]{2025arXiv250511691K}\footnote{\url{https://astro.phys.ntnu.no/SpiderCAT}}.

Due to the close proximity of the pulsar and its companion, the relativistic pulsar wind can ablate material from its companion star, gradually evaporating it over time \citep{1988Natur.333..237F}. The ablated material forms an ionised medium, which can then be seen at radio wavelengths as long-duration frequency-dependent eclipses \citep[e.g.,][]{Shang2024, 2024MNRAS.532.4089A, 2024ApJ...973...19K, Polzin2018}. The exact physical mechanism causing these eclipses remains uncertain, though \cite{1994ApJ...422..304T} has proposed a number. These mechanisms can be investigated by measuring changes in the dispersion measure, $\text{DM} = \int_0^d n_e(l) \, dl$, near the eclipse boundaries of spider systems, using radio time--domain observations. Here, $n_e(l)$ represents the number density of free electrons (in $\mathrm{cm}^{-3}$) at a given point $l$ along the line of sight from the pulsar (at distance $d$) to the observer (at $l=0$). An increase in DM at eclipse ingress or egress indicates the presence of additional ionised material between the pulsar and the observer, providing insight into the plasma environment surrounding the companion. This approach has been employed in several studies aimed at identifying the underlying eclipse mechanism. For example, observations of PSR\,J1908+2105 \citep{2025ApJ...982..168G} and PSR\,J1431$-$4715 \citep{2024ApJ...973...19K} suggest synchrotron absorption as the favoured eclipse mechanism, while PSR\,J1227$-$4853 \citep{Kudale2020} shows evidence consistent with cyclotron absorption.

The long-duration eclipses of spider pulsars, as well as their changing accelerations due to their short orbital periods, cause their apparent spin frequencies to vary significantly throughout observations, making the identification of spider systems in the time domain challenging. With the development of increasingly sensitive and wider-bandwidth low-frequency radio instruments such as the SKAO telescopes and their precursors \citep{Dewdney2009}, image domain searches are becoming increasingly viable for detecting such systems (e.g., \citealt{2024MNRAS.528.5730Z, 2025PASA...42..139P}). These systems can then be followed up with targeted beamforming observations. 

Pulsar timing is a powerful tool in many areas of astrophysics, with applications ranging from testing general relativity to the detection of gravitational waves \citep[e.g.,][]{2023MNRAS.519.3976M,2006Sci...314...97K}. The intrinsic and orbital properties of spider pulsars can be determined through timing studies, such as the pulsar’s spin period, its period derivatives, and the orbital period. The timing of spider pulsars involves measuring the time of arrival (ToA) of the pulses detected at a telescope and comparing them to a timing model. This can provide insight into the nature of the system’s eclipsing material by analysing the delayed ToAs just before and after the pulsar enters and exits eclipse. In addition, many spider pulsars exhibit orbital period variations, which can be tracked through long-term timing campaigns \citep{2025ApJ...982..170R}. These variations are thought to be driven by processes such as changes in the gravitational quadrupole moment of the companion star, which extracts orbital energy from the system \citep{2020MNRAS.494.4448V}. 

Spider MSPs are also seen to emit $\gamma$-ray emission \citep[e.g.,][]{2025ApJ...978..106L, 2024MNRAS.530.4676T,2021ApJ...909....6D}. The Third Pulsar Catalogue (3PC) \citep{2023ApJ...958..191S}, compiled from observations with the Large Area Telescope (LAT) onboard the \textit{Fermi}-gamma ray telescope \citep[]{2009ApJ...697.1071A, 2013ApJS..208...17A}, currently contains 294 detected $\gamma$-ray pulsars, including 32 BWs and 13 RBs \citep[]{2023ApJ...958..191S}. Pulsars represent the largest Galactic class of $\gamma$-ray sources. They are characterised by their low $\gamma$-ray variability and high spectral curvature, which sets them apart from blazars (the most numerous extragalactic class). The discovery of additional $\gamma$-ray sources contributes to the identification and classification of previously unassociated \textit{Fermi} sources.

The detection and characterisation of spider pulsars provides valuable insight into pulsar evolution, the equation of state of dense matter, and the study of plasma physics. This paper reports the discovery and follow-up of a new RB spider pulsar discovered in the image domain. The details of the discovery and follow-up observations are outlined in Section~\ref{sec:obs}, along with potential \textit{Fermi} and optical counterparts. The timing methodology is described in Section~\ref{sec:timing_methods}, and the results are presented in Section~\ref{sec:results}. Section~\ref{sec:disscussion} discusses the system parameters and eclipse mechanism. Finally, we conclude our findings in Section~\ref{sec:conclution}.

\section{Observations} \label{sec:obs}

In this section, we outline the discovery and follow-up radio observations of \psr. Observations were carried out using multiple radio telescopes across a range of frequencies, using both beamforming and interferometric imaging modes.

\subsection{Discovery With ASKAP}

Two surveys carried out using Australian Square Kilometre Array Pathfinder \citep[ASKAP]{2021PASA...38....9H} are the Evolutionary Map of the Universe (EMU) \citep{Norris2011PASA...28..215N,Norris2021PASA...38...46N} and the Variables and Slow Transients Survey (VAST) \citep[][]{Murphy2013PASA...30....6M,Murphy2021PASA...38...54M}.

\begin{figure}[t]
    \includegraphics[width=\linewidth]{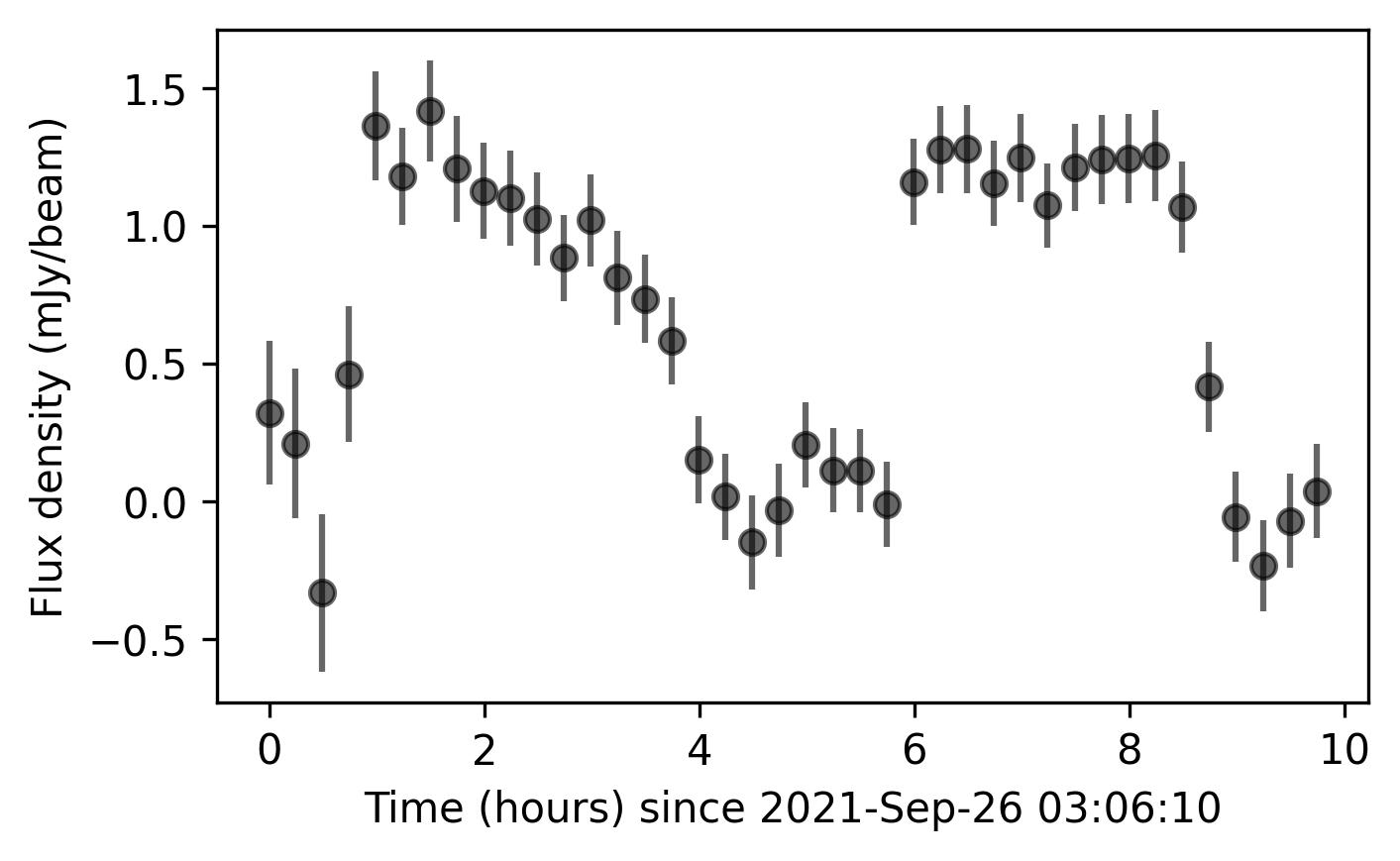}
    \caption{Radio light curve of \source\ in the ASKAP observation SB32526 at 15-min time resolution.}
    \label{fig:figure_J1728_radio_lc}
\end{figure}

\source\ was discovered in the ASKAP archival dataset SB32526 as part of the EMU survey, observed for 10 hours at a central frequency of 943.5\,MHz on 2021~September~26. An untargeted 15-minute timescale search for variable and transient sources using the VASTER pipeline \citep{Wang2023MNRAS.523.5661W} revealed an eclipse-like light curve with a $\sim$5\,h period (Figure~\ref{fig:figure_J1728_radio_lc}). A second 10-hour EMU observation (SB53300), processed in the same way, showed a similar light curve shape and period, confirming the variability.

\begin{figure}[t]
    \centering
    \includegraphics[width=\linewidth]{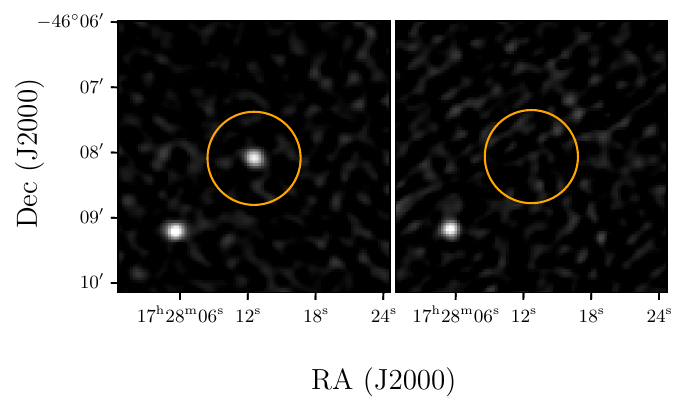}
    \caption{\source\ images from VAST. \textit{Left:} Image from 2023-05-21T16:15:06.6 showing \psr\ is ``on''. \textit{Right:} Image from 2023-09-25T10:23:47.4 showing \psr\ is ``off''.}
    \label{fig:vast}
\end{figure}

\source\ has also been detected as a highly variable source in VAST (see Figure~\ref{fig:vast}), a radio survey designed to detect transient sources in the image domain. VAST is a multi-epoch, short (15-minute) survey, with the Galactic component conducted on a bi-weekly cadence. Observations of \source\ span from 19~November~2022 to 3~October~2024, at a central frequency of 888\,MHz, with a typical rms sensitivity of 0.24\,mJy beam$^{-1}$. 

%\source\ has also been detected as a highly variable source in the ASKAP Variables and Slow Transients (VAST) survey (a multi-epoch short 15-min survey at $\sim$weeks cadence; \citealt{Murphy2013PASA...30....6M,Murphy2021PASA...38...54M}). These light curves are shown in \todo{Figure xx}. 

%To estimate the spectral index, we selected and time-averaged data during the ``on'' state, extracted the full 288\,MHz bandwidth data, and performed a power-law fitting on frequency domain at a 16\,MHz resolution. The in-band fitting gives a steep spectral index $\alpha\sim-2.9\pm0.4$. 

\subsection{Radio Follow-up}

\subsubsection{ATCA}
We observed \source\ at 2100\,MHz using the Australia Telescope Compact Array (ATCA) on 2023~November~18 from 02:30:00 to 07:30:00 UT (project code: C3363). 
The array observed in a compact configuration, H214, and therefore, we removed all short baselines (i.e., the baselines without antenna CA06) during imaging. We measured a flux density of $0.29\pm0.05$\,mJy at 2100\,MHz with a bandwidth of 2000\,MHz. 

%As a result, the sensitivity of the observation was reduced. 
% While we measured a continuum flux density of \todo{xxx} at 2100\,MHz with a bandwidth of 2000\,MHz, no pulse detections were recorded. 
%The ATCA observation gives a better location 

\subsubsection{Parkes/Murriyang}\label{sec:parkes}
\begin{figure}[t]
    \centering
    \includegraphics[width=\linewidth]{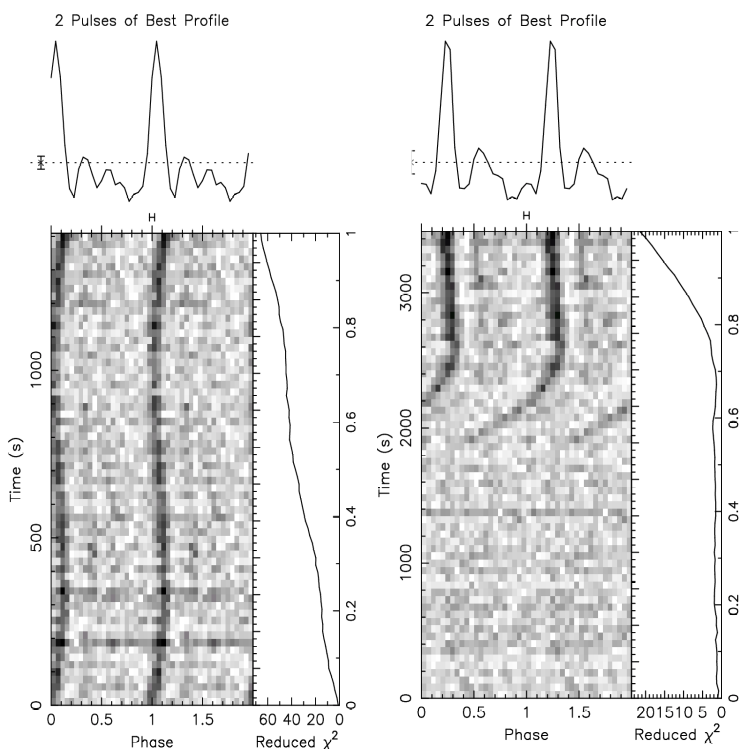}
    \caption{Pulsation detection of \psr\ with Parkes radio telescope on 17~November~2023 (see Section~\ref{sec:parkes} for observation details). 
    The main panel shows the evolution of pulsations over time, folded on the spin period of 2.86\,ms. The top sub-panel displays the integrated pulse profile (frequency-scrunched to enhance signal-to-noise), while the side panel shows the reduced \(\chi^2\) as a function of time, indicating the significance of the detection. 
    \textit{Left:} The pulsar is detected out of eclipse; the curvature of the signal indicates orbital acceleration due to the presence of a companion star.
    \textit{Right:} The pulsar signal disappears as it enters eclipse.}
    \label{fig:pulsar_images}
\end{figure}

We observed \source{} with the Parkes telescope on 2023~November~17 from 06:00:00 to 08:30:00 UT (project code: PX118). 
The observation covered a frequency range of 704--4032 MHz \citep[][]{2020PASA...37...12H}, divided into 3328 channels with 1\,MHz bandwidth each. Coherent de-dispersion was applied at a dispersion measure (DM) of 90\,pc\,cm$^{-3}$, based on the estimated DM from the YMW16 Galactic electron density model \citep{Yao2017ApJ...835...29Y}, using the source position and a initially estimated distance inferred from the parallax of its Gaia companion (see Section~\ref{subsec:gaia}).
To search for pulsations, we used the \textsc{presto} software suite \citep{Ransom2002AJ....124.1788R}, applying a Fourier drift-rate search over a $z$-range of $\pm 150$ \citep{Andersen2018ApJ...863L..13A} to ensure sensitivity to short-period binary pulsars \citep{Ng15}. 
The search was conducted over a DM range of 40--140 pc cm$^{-3}$ using the 960--3008 MHz sub-band, which was cleaned of radio frequency interference (RFI). 
We detected a 2.86\,ms pulsar-like signal at DM of 65.6\,pc\,cm$^{-3}$, accompanied by a measurable acceleration---indicative of binary motion, likely within an eclipsing binary system (See Figure~\ref{fig:pulsar_images}). Hereafter, we refer to \source{} as \psr{}.

Subsequent observations of \psr{} were taken under project code P1342, again using the UWL band (704--4032 MHz). A total of five observations were taken between January to May 2025, with durations of approximately 4--5\,hours. 

%These observations of \psr{} are used to time the pulsar, to precisely determine the intrinsic properties of the pulsar as well as the orbital parameters of the system.
%Details of the timing methodology will be described in Section~\ref{sec:timing_methods}.

\subsubsection{uGMRT}
\psr{} was observed with the upgraded Giant Meterwave Radio Telescope \citep[uGMRT;][]{1991CSci...60...95S, 2017CSci..113..707G}, using band-4 (550--750\,MHz) with a 200\,MHz bandwidth. Simultaneous observations were carried out in both interferometric and coherent phased array modes, using 81-$\mu$s sampling. The observation was taken on the 12\textsuperscript{th}~July~2024 under project code 46\_056, and spanned the whole orbital period ($\sim5$\,hrs). 

The observation began with a scan of the bright calibrator 3C286 for setup, flux, and bandpass calibration. Phasing and phase calibration scans were then performed using 1830–360. \psr{} was observed in 35-minute intervals, interleaved with regular phasing and phase calibration scans to maintain coherence for the phased array mode and to calibrate the gains of the interferometric visibilities.

Calibration and flagging on the interferometric data were carried out using Common Astronomy Software Applications \citep[CASA;][]{2022PASP..134k4501C}. Imaging was done using \textsc{WSClean} \citep{2014ascl.soft08023O}, using multi-frequency synthesis imaging, Briggs weighting with a robust parameter of $-$0.5 \citep[]{1995AAS...18711202B}, and joint-channel deconvolution. The bandwidth was split into four channels, and the data were divided into ten time intervals during imaging. \textsc{Aegean} \citep[]{2012ascl.soft12009H, 2018PASA...35...11H} was used to extract the flux densities and associated errors. When the source is eclipsed, the flux density is reported as an upper limit by measuring the pixel value at the source's location.

The beamformed data were cleaned and converted into SIGPROC filterbank format using \textsc{RFIClean} \citep{2021A&A...650A..80M}, producing a data format suitable for the timing analysis described in Section~\ref{sec:timing_methods}.

\subsubsection{MeerKAT}
MeerKAT S-band observations were conducted under proposal code SCI-20241101-NH-01, using the S4 window at 2625--3500\,MHz. \psr{} was observed in two 30-minute integrations on 15-Dec-2024 at 05:36:39.0 and 09:47:03.6. These observations were scheduled to coincide with the pulsar’s inferior conjunctions (i.e., least likely to be in eclipse). 

The observing setup included standard calibrations: bandpass, polarisation, and phase calibrators, as well as phase-up and test pulsar observations. In addition to correlator observations undertaken at 8-s/854.492-kHz resolution, we also employed the Pulsar Timing User Supplied Equipment \citep[PTUSE;][]{2020PASA...37...28B} in search mode, using 37.45-$\mu$s sampling.

Correlator data were calibrated using the standard SARAO SDP calibration pipeline, and imaged using \textsc{WSClean}. Each 30-minute scan was imaged separately to optimise signal-to-noise for accurate source position measurements and to provide some temporal resolution for studying source variability at S-band. 

\psr's position was obtained using the MeerKAT imaging data, which provides the highest resolution among the available observations. It was measured using \textsc{Aegean} and found to be 17:28:12.27, $-46$:08:01.26 (J2000) with uncertainties of 42\,mas in RA and 28\,mas in Dec (1$\sigma$).

%The PTUSE data is used in the timing analysis, which is detailed in Section~\ref{sec:timing_methods}.

\subsection{Multiwavelength Archival Data And Follow-up}

This section details the optical and $\gamma$-ray archival searches for counterparts of \psr.

\subsubsection{Fermi}\label{sec:Fermi}

\begin{figure}[t]
    \centering
    \includegraphics[width=\linewidth]{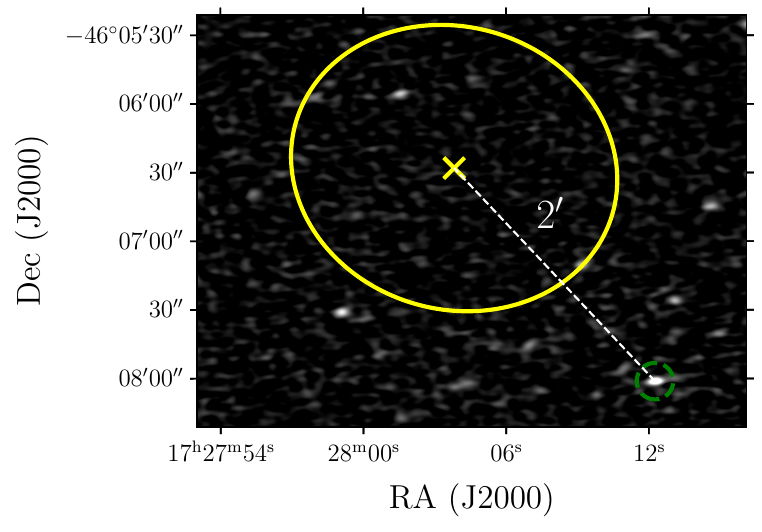}
    \caption{MeerKAT S-band image marking the location of the \psr\ (green circle) and the closest \textit{Fermi} source 4FGL J1728.0-4606 (yellow cross). The position offset and \textit{Fermi} 95\,$\%$ error ellipses (shown in yellow) are indicated. See Section~\ref{sec:Fermi} for more details.}
    \label{fig:fermi-sep}
\end{figure}

We searched for possible $\gamma$-ray associations of \psr\ in \textit{Fermi's} Fourth Full Catalogue of LAT Sources \citep[4FGL-DR4;][]{2020ApJS..247...33A}. The nearest source is \fermi, which lies $2'$ away from \psr, as shown in  Figure~\ref{fig:fermi-sep}. Our source appears to lie outside the 95\% confidence ellipse. While this may indicate a low probability of association, \citet{2020ApJS..247...33A} notes that $\gamma$-ray source localisation in the Galactic Plane is challenging due to high background emission. Therefore, we still consider this a potential association.

%A detection of $\gamma$-ray pulsations using the radio timing ephemeris would provide a definitive test of this association; however, this lies beyond the scope of the present work.

\subsubsection{Gaia}
\label{subsec:gaia}

We searched archival data for a possible companion star to \psr. A nearby star, \gaia{}, was identified in Gaia Data Release 3 \citep{2023A&A...674A...1G}, classified as an eclipsing binary with a similar orbital period to that observed in the radio lightcurve of \psr. Hereafter, we refer to \gaia{} as \gaiaone{}.

The position of \gaiaone{} is 17:28:12.1689, $-$46:08:00.8634 (J2016), with positional uncertainties of  0.2077\,mas in RA and 0.1394\,mas in DEC.

\begin{figure}[t]
    \centering
    \includegraphics[width=\linewidth]{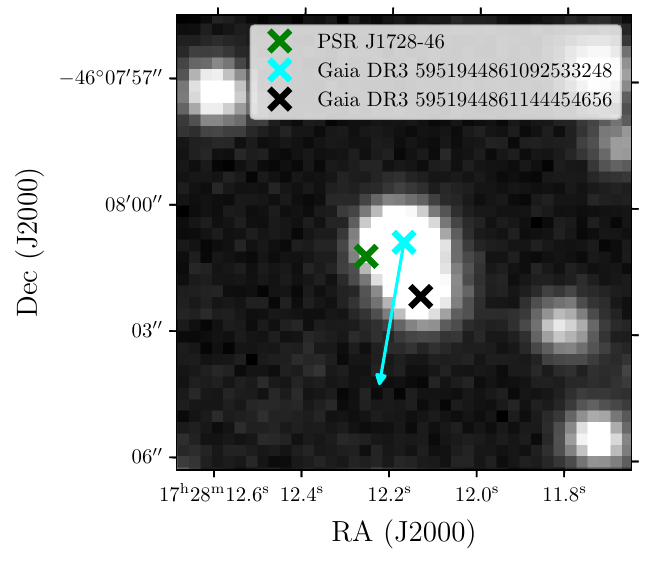}
    \caption{The image shows the position of \psr{} relative to its potential optical companion \gaiaone{}. The background image was obtained with the DECam instrument on the CTIO 4-m Blanco telescope in the VR-band filter (central wavelength 630\,nm, bandwidth 260\,nm) as part of the 2016 observations \citep{2015AJ....150..150F}. The source is a blend of two objects, the second being Gaia DR3 5951944861144454656. The cyan arrow indicates the proper motion vector of the \gaiaone{} source, with components $(\mu_{\alpha} \cos\delta, \mu_\delta) = (+2.094, -8.844)$~mas~yr$^{-1}$. The positional uncertainties of the Gaia sources and the pulsar are on the order of milliarcseconds and are not shown in the figure.}
    \label{fig:gaia}
\end{figure}

The positional offset between \psr\ and \gaiaone{} is less than $0.5''$ (see Figure~\ref{fig:gaia}). \gaiaone{} has a mean G-band magnitude of 18.8\,mag and parallax of 0.63$\pm$0.23\,mas. We present the light curve in Section~\ref{sec:optical_results}, confirming the association.

%Gaia is a key mission in the science programme of the European Space Agency \citep[ESA][]{2013IAUS..291..127R}

\subsubsection{Las Cumbres Observatory}

We conducted 17.8 hours of optical imaging of \gaiaone{} with the \textit{Sinistro} cameras mounted on 1m telescopes as part of the Las Cumbres Observatory Global Telescope Network \citep[LCOGT;][]{2013PASP..125.1031B} between May and July 2024.  This consisted of $g$, $r$, $i$ and $z$-band observations across the orbital period with exposure times of 140, 100, 80 and 160\,s respectively, which were automatically reduced with the BANZAI pipeline \citep{2018SPIE10707E..0KM}.  For each filter, the resultant images are stacked with \textsc{SWarp} \citep{2010ascl.soft10068B} in groups of three consecutive images to minimise the impact of cosmic rays and maximise sensitivity while maintaining an acceptable cadence.  Aperture photometry is extracted  using \textsc{Source-Extractor} \citep{1996A&AS..117..393B} and zeropoints are measured using the SkyMapper Sky Survey DR4 \citep{2024PASA...41...61O}. It is also important to note that \gaiaone{} is blended with another \textit{Gaia} source (\textit{Gaia} DR3 5951944861144454656, located 1.36$''$ from \gaiaone{}) in the LCO images. \textsc{Source-Extractor} was used to force the de-blending of the two sources.

\begin{table*}
    \centering
    \begin{tabular}{c|c|c|c|c|c}
    \hline
    Observatory & Num. Observations & Num. ToAs & Frequency (MHz) & Span & Project Code \\
    \hline
    uGMRT & 1 & 265 & 550--750 & 12~Jul~2024 & 46\textunderscore056\\
    Parkes & 7 & 560 & 704--4032 & 17~Nov~2023--5~May~2025 & PX118, P1342  \\
    MeerKAT & 2 & 273 & 2625--3500 & 15~Dec~2024 & SCI-20241101-NH-0 \\
    \hline
    \end{tabular}
    \caption{Summary of timing observations from each Telescope.}
    \label{tab:timing_details}
\end{table*}

\section{Pulsar Timing Methodology}\label{sec:timing_methods}

Pulsar timing involves a repetitive process of fitting a model for the intrinsic properties of the pulsar (including its astrometric, rotational, and binary parameters) to the pulse arrival times, eventually yielding a coherent timing solution that accounts for every rotation of the pulsar over the observation span. This section describes the methodology used to obtain a coherent timing solution for \psr. To do this, we use the beamforming data from uGMRT (550--750\,MHz), Parkes (704--4032\,MHz), and MeerKAT (2625--3500\,MHz), which are detailed in Section~\ref{sec:obs}. 

\subsection{Initial Orbital Solution}

To carry out the timing analysis of \psr{}, we need an initial estimate of the system's intrinsic parameters. Two of the observations obtained with Parkes (under project code P1342) were taken within a week of each other and covered the full orbital phase ($\sim$5\,hours), allowing us to constrain the orbit. We used \textsc{fitorbit.py}\footnote{\url{https://github.com/emmanuelfonseca/PSRpy}}to fit an orbital model and obtain initial estimates of the pulsar’s spin period, orbital period, projected semi-major axis, epoch of periastron, eccentricity, and argument of periastron.

\subsection{Pulse Times-of-Arrival}

We folded the data using the initial ephemeris, selecting the ELL1 binary model \citep{Lange2001}, which is suitable for systems with very small eccentricities. The data were folded using \textsc{dspsr}, part of the \textsc{PSRCHIVE} package \citep{2012AR&T....9..237V}, using 8 CPU threads and with the data integrated into 10-second intervals. The data were subsequently binned in frequency and time using \textsc{pam}. For MeerKAT, the data were binned into 8 frequency channels and $\sim$1\,minute sub-integrations; for Parkes, four frequency channels and $\sim$5\,minute sub-integrations were used; and for uGMRT, 8 frequency channels and $\sim$3\,minute sub-integrations. 

We then used \textsc{pat} to extract ToAs by determining the phase shift between the high S/N pulsar template and the observed profiles, using a Fourier-domain Markov Chain Monte Carlo algorithm. We constructed a high S/N template for each telescope (uGMRT, Parkes, and MeerKAT) using \textsc{paas}, by phase summing the highest S/N profiles from the respective observations. For each telescope, we use a single pulse template across the entire observing band, assuming the intrinsic pulse shape evolves negligibly across frequency. As shown in Section~\ref{sec:pulse_profile}, the primary frequency-dependent change in the profile is a slight widening of the pulse, which we attribute to scattering broadening. According to \citet{2025MNRAS.tmp.1156M}, a scattering timescale of $91 \pm 34$\,{\textmu}s at 704\,MHz (the lowest frequency of the Parkes observations) implies an additional ToA error of approximately 45\,{\textmu}s. This is much smaller than the residual errors ($\sim$300\,{\textmu}s) we observe, and thus, using a single template across the band should not introduce significant bias. The specific number of ToAs from each observation is determined by observation length and the brightness of the detection. Table~\ref{tab:timing_details} gives details of the number of ToAs, where we have excluded ToAs with uncertainties $>150\,\mu\text{s}$ and ToAs in the eclipse region.

%, corresponding to phases $\sim$0.15--0.4.

\subsection{Phase Connected Timing Solution}

The ToA residuals were plotted using \textsc{Tempo2} \citep{2006ChJAS...6b.189H}, where residuals refer to the difference between the observed ToAs and those predicted by the timing model. To achieve a phase-connected solution across all observations, we employed an iterative approach in which observations were added incrementally, and model parameters were refined accordingly. We began by fitting only for the spin frequency and binary parameters. As additional observations were incorporated, the model was extended to include the spin frequency and orbital period derivatives, as well as the DM. To account for arbitrary time offsets between telescopes, we fitted \textsc{JUMPs} to the uGMRT and MeerKAT data using \textsc{Tempo2}, with the Parkes data serving as the reference. The JUMPs account for differences such as instrumental delays, the use of distinct pulse templates, and the choice of the fiducial point on each template. After incorporating all datasets, the fit converged on a coherent timing solution.

\section{Results} \label{sec:results}
\subsection{Timing Solution}

\begin{table*}[h]
    \centering
  
    \begin{tabular}{lc}  
    \hline
    \textbf{Observation and data-set parameters} & \\
    \hline
    Pulsar name ................................................................................................... & \psr{} \\ 
    MJD range ......................................................................................................& 60797.7$--$60265.1  \\
    Total time span (year) ...........................................................................................& 1.5 \\
    Number of ToAs ................................................................................................ & 1098 \\
    Epoch of frequency determination (MJD) .........................................................................& 60686 \\
    Epoch of position determination (MJD) ..........................................................................&  60686\\
    Epoch of dispersion measure determination (MJD) ................................................................& 60686  \\
    Reduced Chi-square, $\chi^2$ ............................................................................................ & 1.9182\\
    Post-fit residual rms ($\mu$s) .................................................................................&  49.667  \\
    \hline
    \textbf{Measured Quantities} &  \\
    \hline
    Right ascension (J2000), $\alpha$ ...............................................................................& 17:28:12.27  \\
    Declinaton (J2000), $\delta$ ....................................................................................& $-$46:08:01.26    \\
    Dispersion Measure, DM (pc cm$^{-3}$) ............................................................................& 65.4856(6) \\
    Spin frequency, $\nu$ (Hz) ......................................................................................& 349.1603451421(1)  \\
    Spin frequency derivative, $\dot\nu$ (Hz s$^{-1}$) ..............................................................& -9.56(1)$\times 10^{-16}$   \\
    Orbital period, $P_\mathrm{orb}$ (days) ....................................................................................& 0.210410559(6)\\
    Orbital period derivative, $\dot P_\mathrm{orb}$  ...........................................................& 3.35(9)$\times 10^{-10}$ \\
    Projected semimajor axis, $x$ (lt-s) ............................................................................& 0.328022(9) \\
    Time of ascending node, $T_{asc}$ (MJD) ..........................................................................& 61291.344981(9) \\
    EPS1, $\epsilon \sin(\omega)$\footnote{$\omega$ is the argument of periastron.} ............................................................................& 0.00016(3)\\
    EPS2, $\epsilon\ \text{cos}(\omega)$ ............................................................................& 0.00001(2)\\
    \hline
    \textbf{Derived Quantities} &  \\
    \hline
    DM distance YMW16 model, $d$ (kpc) ...................................................................................& 2.2 \\
    DM distance NE2001 model, $d$ (kpc) ..................................................................................& 1.8 \\
    Minimum companion mass, $M_c$ ($M_\odot$) .....................................................................................& 0.13 \\
    Spin down energy loss, $\dot E$ ($erg s^{-1}$)........................................................................& 1.11$\times 10^{34}$\\
    Characteristic age, $\tau _c$ (Gyr) ..................................................................................& $\sim$5.8 \\
    Surface magnetic field, $B_0$ (G) ....................................................................................& 1.5$\times 10^{8}$\\
    
    \hline
    \end{tabular}
    \caption{Timing parameters for \psr.}
    \label{tab:timing_results}

\end{table*}

\begin{figure}[t]
    \centering
    \includegraphics[width=\linewidth]{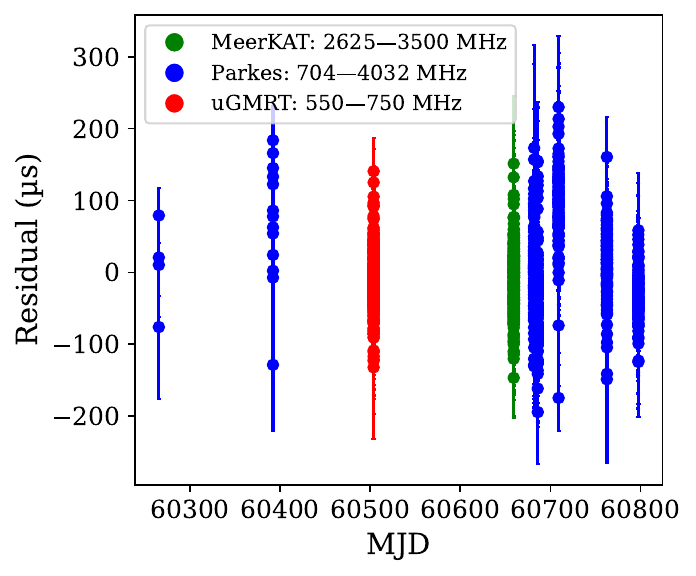}
    \caption{Timing residuals of \psr\ fitted using the ephemeris given in Table~\ref{tab:timing_results}. Data corresponding to the eclipse region (orbital phase $\sim$0.27--0.72, see Figure~\ref{fig:excess_dm}) have been excluded from the fit.}
    \label{fig:residuals}
    \end{figure}

\psr{} is found to be an MSP with a spin period ($P$) of 2.86\,ms, in a 5.05\,hr binary orbit with a companion with a derived minimum mass of 0.13$M_\odot$, assuming an inclination of 90$^\circ$ (for an edge-on orbit) and pulsar mass of $1.4M_\odot$\citep{2011A&A...527A..83Z}. We present the timing solution in Figure~\ref{fig:residuals} and in Table~\ref{tab:timing_results}. The reduced $\chi^2\approx1.9$ is greater than unity and may add additional uncertainty to the timing solution that has not been accounted for. Using our timing parameters in Table~\ref{tab:timing_results}, we derive the spin-down luminosity as $\dot E =1.11\times 10^{34}\text{erg\ s}^{-1}$ using the formula $\dot E\equiv4\pi^2I\dot PP^{-3}$, where the neutron star moment of inertia is assumed to be $I=10^{45}\ \mathrm{g\ cm^2}$ and $\dot P$ has been corrected for the Shklovskii effect (see Section~\ref{sec:period_derivative} for details). The characteristic age and surface magnetic field are calculated as $\tau _c\equiv\frac{P}{2\dot P}=5.8$\,Gyr, and $B_0\equiv3.2\times10^{19}(P\dot P)^{\frac{1}{2}}=1.6\times 10^8$\,G, respectively. These values are also given in Table~\ref{tab:timing_results} and are all consistent with the known MSP population. We also measure a significant detection for the orbital period derivative $\dot P_b$ and explain the implications of this in Section~\ref{sec:period_derivative}.

\subsubsection{Pulse Profile Evolution}\label{sec:pulse_profile}

\begin{figure}[t]
    \centering
    \includegraphics[width=\linewidth]{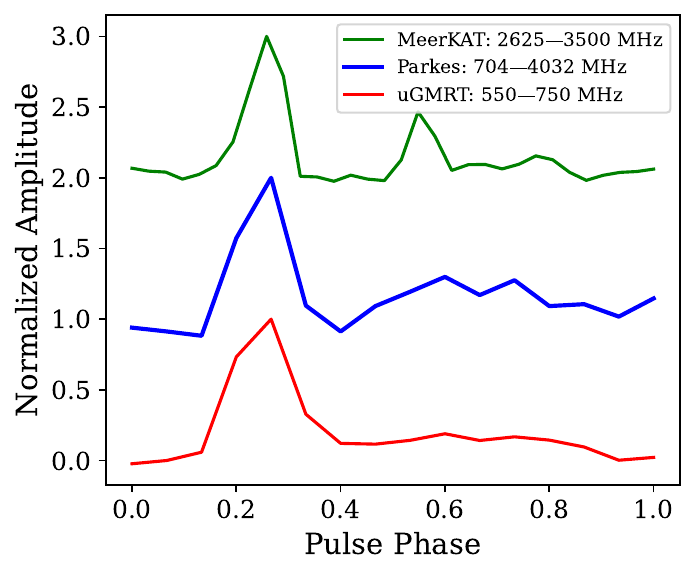}
    \caption{Normalised, averaged pulse profiles outside the eclipse phase at different observing frequencies. A second component is visible in the MeerKAT profile; see Section~\ref{sec:pulse_profile} for details.}
    \label{fig:pulse_profile}
\end{figure}

The frequency-dependent evolution of pulse profiles offers key insights into the emission mechanisms of pulsars, shedding light on both emission geometries and radius-to-frequency mapping \citep[][]{1978ApJ...222.1006C}. \psr{} is seen to exhibit frequency-dependent pulse profile evolution,  with a second component seen at 3062\,MHz as seen in Figure~\ref{fig:pulse_profile}. We fit Gaussian profiles to the dominant and secondary pulse components using \textsc{paas}, deriving pulse widths as a function of frequency.  The main component has a width at 50\% of the peak intensity $W_{50}$ of 0.080\,ms at 650\,MHz, 0.14\,ms at 2368\,MHz and 0.22\,ms at 3062\,MHz. The second component has $W_{50} = 0.44$\,ms at 3062\,MHz. 

Radio waves emitted by pulsars propagate through the turbulent, ionised interstellar medium (ISM), which introduces frequency-dependent delays and phase variations. One important propagation effect is scattering, which causes multipath propagation and temporal broadening of the pulse. This effect is evident in Figure~\ref{fig:pulse_profile}, where increased scattering is observed at lower frequencies. To model this pulse broadening, we assume the thin screen approximation, in which the scattering timescale is determined by fitting the observed pulse profile with a convolution of a Gaussian and an exponential function \citep{1968Natur.218..920S}.  We obtain scattering timescales of $130 \pm 50$\,\textmu s, $0.5 \pm 0.2$\,\textmu s, and $0.14 \pm 0.05$\,\textmu s at 650, 2348, and 3062\,MHz, respectively, where the uncertainty estimates are derived from the covariance matrix of the fit. Using the DM-based model from \citet{2004ApJ...605..759B}, we estimate scattering timescales $11.7791 \pm 0.0004$\,\textmu s, $0.080139 \pm 0.000003$\,\textmu s, and $0.029715 \pm 0.000001$\,\textmu s at the same frequencies, where the uncertainties are derived from error propagation of the DM measurement. The significantly larger scattering timescales derived from the thin screen model indicate the presence of additional scattering beyond what is predicted by the DM-based model. This discrepancy likely arises from local plasma within the binary system, such as material from the companion star, which is not accounted for in the average ISM scattering relations.

\subsubsection{Excess DM}\label{sec:excess_dm}

\psr{} is eclipsed for approximately 45\% of its orbit at 2368\,MHz. As the pulsar is eclipsed, there is an excess in the observed DM due to the presence of the eclipsing material from the ablated companion. 

The excess DM (DM$_\text{excess}$) is given by the equation \citep{2004hpa..book.....L}

\begin{equation}
    \mathrm{DM_{excess}\ (pc\ cm^{-3})} = 2.4\times10^{-10}\ t(\mu s) \ \nu(\text{MHz})^{2} \ ,
\end{equation}\label{eq:excess_dm}

\noindent where $t$ is the ToA time delay and $\nu$ is the frequency. The corresponding electron column density ($N_{e}$) can be obtained using

\begin{equation}
    N_{e}\ (\text{cm}^{-2}) = 3\times10^{18} \times \mathrm{DM_{excess}\ (pc\ cm^{-3})} 
\end{equation}\label{eq:N_e}

\begin{figure*}[t]
    \centering
    \includegraphics[width=\linewidth]{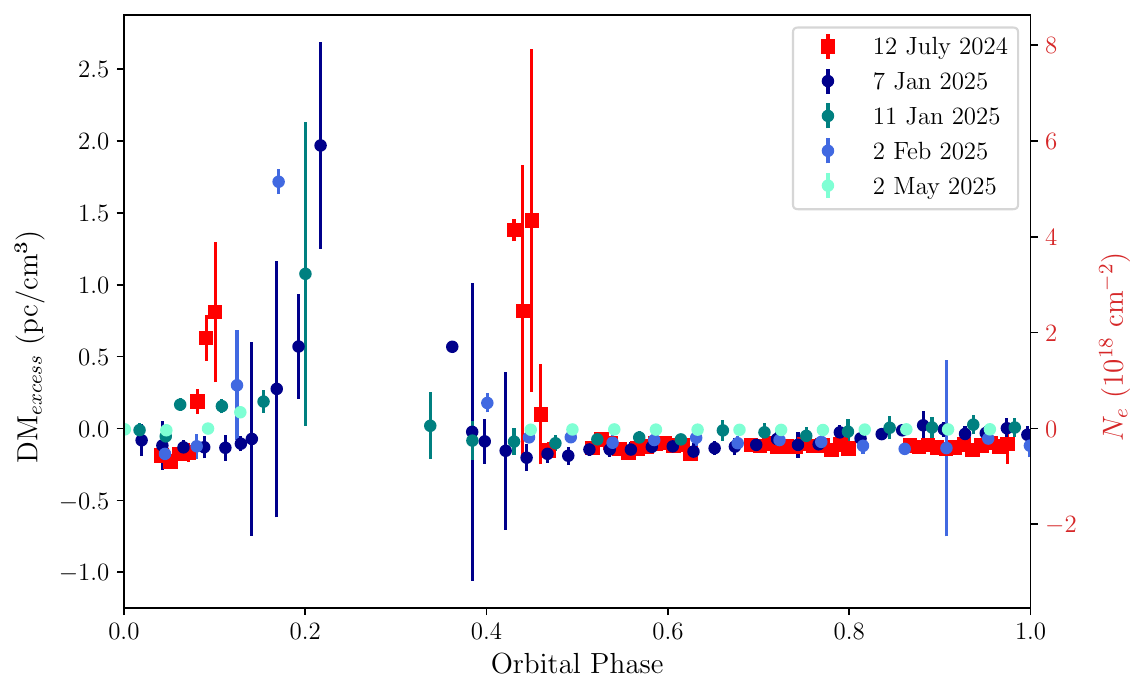}
    \caption{
    DM\textsubscript{excess} (left y-axis) and the corresponding $N_e$ (right y-axis) are shown against orbital phase.  uGMRT data at 650\,MHz are shown as red squares, and Parkes data at 2368\,MHz as circles. The orbital phase is defined with respect to $T_{asc}$, with the eclipse centred at an orbital phase of approximately 0.25. See Section~\ref{sec:excess_dm} for further details.}
    \label{fig:excess_dm}
\end{figure*}

Figure~\ref{fig:excess_dm} shows DM\textsubscript{excess} against orbital phase, where the points in the eclipsed phase have been removed. The orbital phase has been defined with respect to the time of ascending node ($T_{asc}=61291.344981(9)$), which is provided in Table~\ref{tab:timing_details}. The figure includes data from the observing epoch that covers the ingress/egress phase and has sufficient S/N. The figure shows an increase in DM in each epoch, with a larger DM excess seen in ingress. Lower frequencies (uGMRT at 650\,MHz) move into eclipse earlier, as there is an increase in DM earlier. The largest DM\textsubscript{excess}$=2.0 \pm 1.2 \ \mathrm{pc\ cm^{-3}}$, which corresponds to $N_e =5.9 \pm 3.6 \times10^{18} \ \mathrm{cm^{-2}}$.

%The ingress, fully eclipsed, and egress phases are shaded in orange, grey, and green, respectively. 
%The orange shaded region denotes the eclipse duration at 888\,MHz ($t_{\mathrm{eclipse},888,\mathrm{MHz}}$; see Section~\ref{sec:imaging_results_radio}).

\subsection{Imaging}\label{sec:imaging_results}

\subsubsection{Radio}\label{sec:imaging_results_radio}
\begin{figure}[t]
    \centering
    \includegraphics[width=\linewidth]{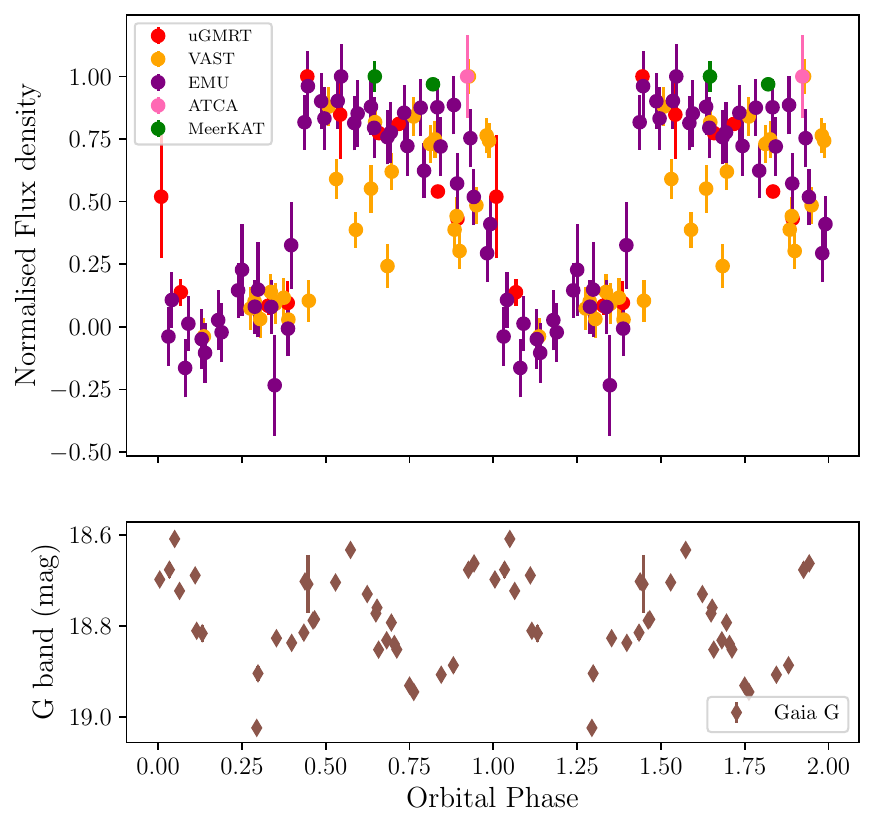}
    \caption{\textit{Top panel:} Radio folded lightcurve of \psr{} on the orbital period of 5.05\,h, obtained through timing analysis. The flux density has been normalised to its maximum. See Section~\ref{sec:imaging_results} for more details. 
    \textit{Bottom panel:} \gaiaone{} G-band optical light curve folded on the same orbital period. The orbital phase is computed with respect to $T_{\rm asc}$ and shown over two consecutive cycles. See Section~\ref{sec:optical_results} for more details.
    }

    \label{fig:folded_lightcurve}
\end{figure}

\psr{} was observed across a range of radio frequencies using multiple radio telescopes in the image domain (see Section~\ref{sec:obs}). The top panel of Figure~\ref{fig:folded_lightcurve} presents the folded light curve on the orbital period derived from our timing analysis. The folded light curve exhibits eclipsing behaviour consistent with that seen in the beamformed observations.

To quantify the eclipse properties, we fit the EMU data (as it covers the full orbital phase) with a double Fermi-Dirac function \citep[see][]{2024MNRAS.528.5730Z}:
\begin{equation}\label{eq:fermi-dirac}
S(\phi) = S_0 \left[ \left( \frac{1}{1 + e^{\frac{\phi - \phi_i}{w_i}}} - \frac{1}{1 + e^{\frac{\phi - \phi_e}{w_e}}} \right) \right]
\end{equation}
where $\phi$ is the orbital phase, $\phi_i$ and $\phi_e$ are the phases at the eclipse ingress and egress, $w_i$ and $w_e$ are the widths of the ingress and egress, and $S_0$ is the flux density at the pulsar's inferior conjunction (i.e., out of eclipse). The fitted parameters $\phi_i$, $w_i$, $\phi_e$ and $w_e$ are found to be 0.98(2), 0.05(2), 0.4(1) and 0.001(30), respectively, with uncertainties derived from the covariance matrix of the fit. The total eclipse duration for EMU, calculated as $(1 - \phi_i) + \phi_e$, is found to be approximately 42\% of the orbital phase. 

For the uGMRT data, which under-samples the egress side of the eclipse, we fit a modified version of Equation~\ref{eq:fermi-dirac}, in which we fit only the ingress side: 
\begin{equation}\label{eq:fermi_dirac_half}
    S(\phi) = S_0 \left[\left( \frac{1}{1 + e^{\frac{\phi - \phi_i}{w_i}}} \right) \right]
\end{equation}
The resulting best-fit parameters for the uGMRT ingress are $\phi_i = 0.848(6)$ and $w_i = 0.091(6)$.

To characterise the frequency dependence, we measure the ingress duration ($t_\text{ingress}$) at 650\,MHz (uGMRT) and 888\,MHz (EMU). We define the ingress duration as the time taken for the flux density to decrease from 90\% to 10\% of its maximum value ($S_0$ in Equation~\ref{eq:fermi-dirac} and \ref{eq:fermi_dirac_half}).  We found $t_{\text{ingress,888MHz}}=1.03$\,hr and $t_{\text{ingress,650MHz}}=1.77$\,hr. Assuming a power-law dependence of ingress duration with frequency ($t_{\text{ingress}} \propto \nu^n$), we estimate a power law index $n=-1.74$. 

%To characterise the frequency dependence of the ingress position, we adopt the same approach as \citet{2020MNRAS.494.2948P} and fit a power law $\phi_e(\nu) = \phi_{e,\nu_0} \left( \frac{\nu}{\nu_0} \right)^n$, finding $n=0.46(7)$.

\begin{figure}[t]
    \centering
    \includegraphics[width=\linewidth]{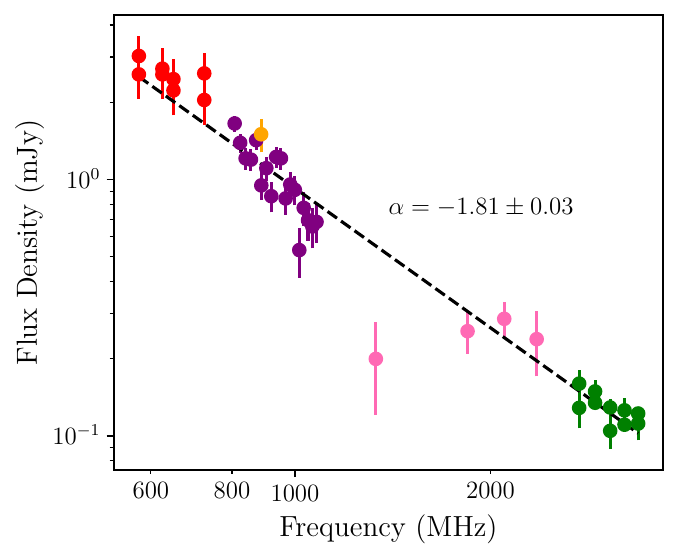}
    \caption{Spectrum of \psr{} fitted with a simple power-law, using flux densities corresponding to the inferior conjunction of the pulsar. The spectral index $\alpha$ is highlighted on the figure. The marker colours are the same as in Figure~\ref{fig:folded_lightcurve}. For more details, see Section~\ref{sec:imaging_results}. }
    \label{fig:spectra}
\end{figure}

Figure~\ref{fig:spectra} illustrates the spectrum fitted with a simple power law of the form $S_\nu \propto \nu^\alpha$, where $\alpha$ is the spectral index. The spectral index found is $-$1.81(3), where the error is derived from the covariance matrix of the fit.

%Continuum imaging is expected to be unaffected by the temporal smearing caused by $DM_{excess}$ and/or scattering -- use the comparative study to understand the eclipse mechanism. The continuum and pulsed flux densities show correlated changes as the pulsar is transitioning into eclipse 

\subsubsection{Optical}\label{sec:optical_results}

Figure~\ref{fig:folded_lightcurve} shows the optical G-band phase-folded lightcurve for \gaiaone{}. The \textit{Gaia} G-band light curve displays two brightness maxima per orbit, consistent with ellipsoidal modulation from a tidally distorted companion. For pure ellipsoidal modulation, the peaks are expected near orbital phases 0 and 0.5, corresponding to the ascending and descending nodes of the companion. However, for our source, the observed maxima occur at phases $\sim0.15$ and $\sim0.6$. This shift is likely caused by irradiation from the pulsar, indicating that the light curve reflects a combination of ellipsoidal and irradiation effects.
 \textit{Gaia} measured an effective temperature ($T_{eff}$) of 4384\,K corresponding to a K-type star. This is consistent with RB systems, which generally have companion temperatures in the range 4000--6000\,K. Gaia DR3 provides a model-dependent distance from the General Stellar Parameterizer from Photometry \citep[GSP-Phot;][]{2023A&A...674A..27A} derived from a Bayesian inference combining the BP/RP spectra, G-band photometry, the measured parallax, and stellar evolutionary models. The calculated GSP-Phot distance is 1.288$^{+0.122}_{-0.091}$\,kpc. \cite{2023A&A...674A..27A} state that the GSP-Phot distance is underestimated for sources with large parallax errors (S/N $<$20). \gaiaone{} has a parallax of 0.63$\pm$0.23\,mas, giving S/N$\sim$2.7, so the GSP-Phot distance is likely unreliable. \cite{2021AJ....161..147B} also provides geometric and photometric distances based on Galactic priors, which are $3.3^{+2.6}_{-1.8}$\,kpc and $4.5^{+1.3}_{-1.0}$\,kpc, respectively. The low parallax S/N causes the priors in both \citet{2021AJ....161..147B} and \citet{2023A&A...674A..27A} to dominate the distance estimates, which can bias the results and likely explains the discrepancy with the DM-based distance of 1.8\,kpc and 2.2\,kpc from the NE2001 and YMW16 electron density models, respectively. Given the unreliability in both the photometric and geometric distance estimates, we do not place much weight on the results. Nonetheless, the orbital period and observed photometric modulations confirm that \gaiaone{} is the companion to \psr{}.

 %This is smaller than the distance inferred from the pulsar's DM, which is 1.8\,kpc and 2.2\,kpc based on the NE2001 and YMW16 electron density models, respectively. This discrepancy may arise from uncertainties in DM-based distances resulting from variations in the Galactic electron density along the line of sight. Nonetheless, the orbital period and observed photometric modulations confirm that \gaiaone{} is the companion to \psr{}.

 %The peaks occur near orbital phases 0.25 and 0.75, corresponding to the ascending and descending nodes, respectively.

We did not obtain reliable results from the LCO photometry. In Section~\ref{sec:optical_results}, we noted that \textsc{SExtractor} was used to deblend our target from its nearby companion. However, given the close separation of the two sources (1.36$''$), this deblending is not fully reliable and introduces significant uncertainties in the photometry. Consequently, we do not attempt any further analysis of these data.

\subsection{Searching For $\gamma$-ray Pulsations}

We searched for $\gamma$-ray pulsations from \fermi{} using Fermi-LAT Pass~8 (P8R3) data \citep{2018arXiv181011394B} in the energy range 0.1–100\,GeV, covering the period from 2023-11-17 to 2025-05-02 (the span of our radio ephemeris). Events were selected within a 2\,$^\circ$ region of interest centred on the position of \fermi{}. No additional cuts were applied at this stage.

Using the radio ephemeris and the Fermi plug-in for \textsc{TEMPO2} \citep{2011ApJS..194...17R}, we calculated the rotational phase for each photon arrival time from \fermi{}. We also split the data into three energy bands: 0.1–0.3\,GeV, 0.3–1\,GeV, and 10–100\,GeV. In all bands, the resulting H-test statistics were below 8, corresponding to significance levels of less than $\sim2\sigma$; the H-test is a statistical test for detecting periodic signals in sparse photon data \citep{1989A&A...221..180D}. We require at least $5\sigma$ detection to confirm pulsations, and thus find no statistically significant evidence in any energy range.

The lack of $\gamma$-ray pulsations could be due to intrinsically low photon counts, high background emission levels in the Galactic plane, an unfavourable beaming geometry, or the limited precision of the current radio ephemeris. The relatively short radio ephemeris ($\sim 1.5$\,yrs) for \psr{} does not determine the spin and orbital parameters precisely enough to coherently fold all LAT photons, which span back to 2008. In a more detailed analysis, standard event-quality and zenith-angle cuts could also be applied to reduce background contamination. A more in-depth search, such as a multidimensional grid over the uncertain parameters \citep[e.g.,][]{2019ApJ...883...42N}, would be required to fully explore the $\gamma$-ray pulsation parameter space, which is beyond the scope of this work.

Nevertheless, we still consider \fermi{} a potential association based on positional coincidence and discuss its properties further in Section~\ref{sec:disc_fermi_prop}.

%Using the radio ephemeris and the Fermi plug-in for \textsc{TEMPO2} \citep{2011ApJS..194...17R}, we calculated the rotational phase for each photon arrival time from \fermi{}. No $\gamma$-ray pulsations were detected, so we cannot confirm that \psr{} is a $\gamma$-ray pulsar. The lack of $\gamma$-ray pulsations could be due to a misaligned beam or because the pulsar is $\gamma$-ray quiet. We note, however, that the source lies in the Galactic plane, where the high level of diffuse $\gamma$-ray background can significantly reduce the signal-to-noise ratio. Nevertheless, we still consider \fermi{} a potential association based on positional coincidence and discuss its properties further in Section~\ref{sec:disc_fermi_prop}.

\section{Discussion} \label{sec:disscussion}
%\subsection{Nature of the source}\label{sec:nature}

\subsection{Orbital Period Variation}\label{sec:period_derivative}

From our timing analysis, we obtained a value of the orbital period derivative, $\dot{P}_\mathrm{orb}= 3.35(9)\times 10^{-10}$. Variations in the orbital period are a characteristic feature observed in many spider pulsars and can arise from several astrophysical mechanisms. To investigate the possible causes of $\dot{P}_\mathrm{orb}$, we follow the approach outlined by \citet{Pletsch2015}, expressing the total change as:

\begin{equation}\label{eq:P_dot}
\dot{P}_\mathrm{orb} = \dot{P}_\mathrm{GW} + \dot{P}_\mathrm{D} + \dot{P}_\mathrm{M} + \dot{P}_\mathrm{Q},
\end{equation}

\noindent where $\dot{P}_\mathrm{GW}$ is the contribution from gravitational-wave emission, $\dot{P}_\mathrm{D}$ arises from Doppler shifts due to system acceleration, $\dot{P}_\mathrm{M}$ is due to mass loss from the binary system, and $\dot{P}_\mathrm{Q}$ reflects variations in the gravitational quadrupole moment of the companion star.

Using Equation (5) from \citet{Pletsch2015}, we calculate the contribution from gravitational-wave emission as $\dot{P}_\mathrm{GW} = -4.6 \times 10^{-14}$, which is four orders of magnitude smaller than the observed value. This suggests that gravitational radiation is unlikely to be the dominant cause of the orbital period variation.

The term $\dot{P}_\mathrm{D}$ in Equation~\ref{eq:P_dot} is further decomposed into three components: the Shklovskii effect $\dot{P}_\mathrm{Shk}$ arising from the system’s proper motion, the Galactic acceleration term $\dot{P}_\mathrm{Gal}$, and acceleration due to binary motion $\dot{P}_\mathrm{acc}$.

Using Equation (7) from \citet{Pletsch2015} and Gaia proper motions of $\mu_{\alpha} = 2.0936\ ~\mathrm{mas\ yr^{-1}}$ and $\mu_{\delta} = -8.8439\ \mathrm{mas\ yr^{-1}}$, we compute the Shklovskii contribution to be $\dot{P}_\mathrm{Shk} = 1.2 \times 10^{-21}$. Assuming any apparent orbital period change due to acceleration is proportional to the spin period derivative, we estimate $\dot{P}_\mathrm{acc} = (\dot{P}/P)\times P_\mathrm{orb} = 5.0 \times 10^{-14}$. The Galactic acceleration term, $\dot{P}_\mathrm{Gal}$, is typically smaller than the other two Doppler-related contributions \citep{2009MNRAS.400..805L}. Therefore, we approximate the overall $\dot{P}_\mathrm{D}$ as $5 \times 10^{-14}$. This value is four orders of magnitude smaller than the observed orbital period derivative, allowing us to rule out Doppler effects as the dominant cause.

The $\dot P_{M}$ term is estimated using Equation (8) in \cite{Pletsch2015}. Assuming the companion star fills its Roche lobe (as calculated in Section~\ref{sec:eclipse_mec}), we obtained a mass loss rate of $\dot M = 9.3\times 10^{-10}$ \,$\text{M}_\odot \text{yr}^{-1}$. This gives $\dot P_{M} = -7.0\times 10^{-13}$. This contribution is therefore not the primary driver of the observed orbital period variation.

\begin{figure}[t]
    \centering
    \includegraphics[width=\linewidth]{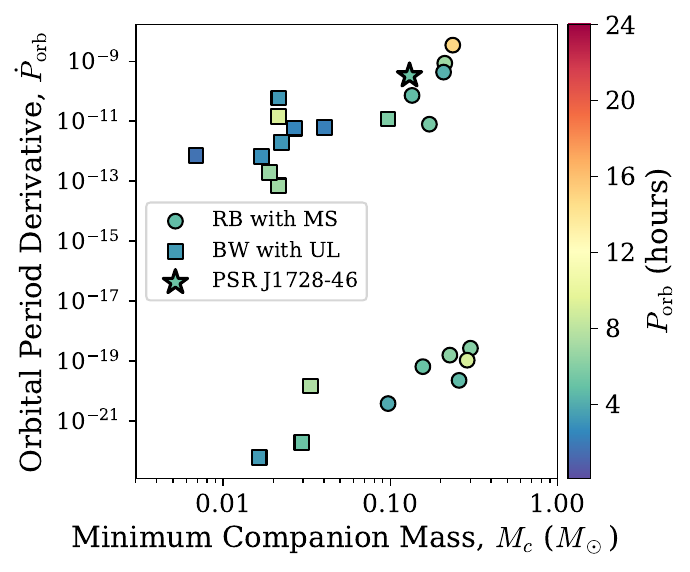}
    \caption{Orbital period derivative ($\dot P_\mathrm{orb}$) against minimum companion mass ($M_c$) for known RB (with main sequence companions) and BW (with ultra-light companions) spider pulsars from the ATNF catalogue, along with \psr{}. The colour map indicated the pulsar's orbital period ($P_\mathrm{orb}$) value. The derived values for \psr{} are given in Table~\ref{tab:timing_results}. }
    \label{fig:period_derivative}
\end{figure}

Therefore, the dominating factor in  $\dot{P}_\mathrm{orb}$ is thought to arise from the gravitational quadrupole moment of the companion \citep{1983A&A...117L...7M}, as discussed for several spider pulsars \citep[e.g.,][]{2020MNRAS.494.4448V, 2011MNRAS.414.3134L}. Figure~\ref{fig:period_derivative} shows $\dot P_b$ against $M_c$ for \psr{}, alongside known spider pulsars with available measurements from the \textsc{ATNF Pulsar Catalogue}\footnote{\url{https://www.atnf.csiro.au/research/pulsar/psrcat/}} (Catalogue Version 2.6.0). Specifically, we include only binary pulsars outside globular clusters with companions classified as ultra-light or main-sequence, $\text{M}_{c}<1\,\text{M}_\odot$ and $P_\mathrm{orb}\leq1$\,day. The figure shows that $\dot P_b$ for \psr{} lies within the parameter space occupied by other RBs.

%Timing the pulsar over longer periods of time may reveal higher-order $\dot P_b$ values. Higher $\dot P_b$ values have been seen in many systems, such as PSR\,J1048+2339 \citep{2016ApJ...823..105D} and PSR\,J2339-0533 \citep{2020ApJ...897...52A}. 

\subsection{Optical Light-Curve Morphology}\label{sec:diss_companion_star}

\begin{figure}[t]
    \centering
    \includegraphics[width=\linewidth]{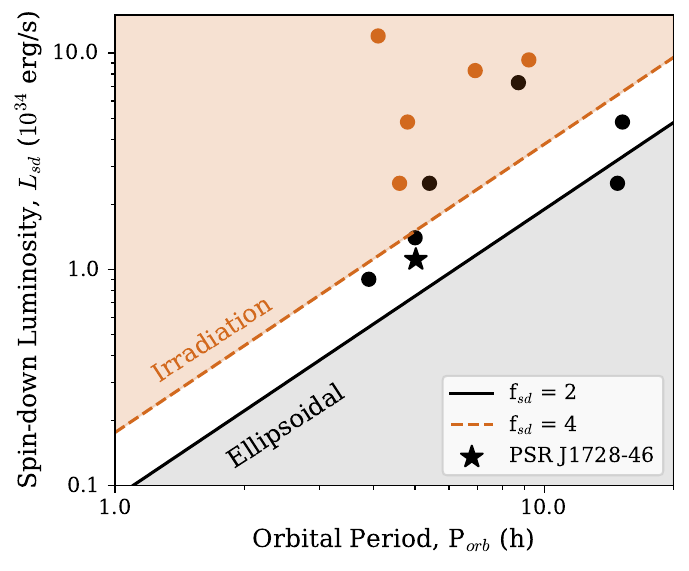}
    \caption{$L_\mathrm{sd}$ versus $P_\mathrm{orb}$ for known RB pulsars, along with \psr{}. This relation $L_\mathrm{sd} \propto P_\mathrm{orb}^{4/3}$ is shown for $f_\mathrm{sd}=2$ (solid line) and $f_\mathrm{sd}=4$ (dashed line) assuming our source parameters $\text{T}_{eff}$= 4384\,K, and $\text{M}_{tot} = 1.5 \text{M}_\odot$. The points are colored according to the number of maxima in the optical light curve per orbit, with black indicating two maxima and orange indicating one. The figure has been adapted from \cite{2023MNRAS.525.2565T}; see Section~\ref{sec:diss_companion_star} for further details.}
    \label{fig:optical_comparison}
\end{figure}
%as approximate upper limit for hotter and more massive RB companions ($T_{eff}$= 5500\,K, $M_{tot} = 2 M_\odot$).

The optical light curves of RB pulsars are either dominated by irradiation or ellipsoidal modulation. Ellipsoidal modulations are observed in about 50\% of known RB systems, where the companion's shape, rather than irradiation, dominates the light curve. Examples include PSR\,J1622$-$0315 \citep[$P_B = 3.9\,h$;][]{2023MNRAS.525.2565T}, PSR\,J1431$-$4715 \citep[$P_B = 10.8$\,h;][]{2019ApJ...872...42S} and PSR\,J1628$-$320 \citep[$P_B = 5.0$\,h;][]{2014ApJ...795..115L}. This difference arises from the interplay of several factors, such as the companion’s intrinsic luminosity, the mass ratio, orbital separation, and the pulsar’s spin-down luminosity.

In Section~\ref{sec:optical_results}, we saw that the \gaiaone{} G-band light curve shows consistency with ellipsoidal modulation. The magnitude of irradiation is quantified through the dimensionless ratio $f_{sd}$ \citep{2023MNRAS.525.2565T}: 

\begin{equation}
    \begin{split}
    f_\mathrm{sd} \simeq 1.1 \times 10^4 
    \left( \frac{L_\mathrm{sd}}{10^{34}\,\mathrm{erg\,s^{-1}}} \right)
    \left( \frac{T_\mathrm{eff}}{10^3\,\mathrm{K}} \right)^{-4} \\
    \times \left( \frac{M_\mathrm{tot}}{M_\odot} \right)^{-2/3}
    \left( \frac{P_\mathrm{orb}}{1\,\mathrm{h}} \right)^{-4/3}
    \end{split}
\end{equation}

where $L_\mathrm{sd}$ is the pulsar spin-down luminosity and $M_\mathrm{tot}$ is the total mass of the system. For our system, we use $L_\mathrm{sd}=1.1\times10^{34}\,\mathrm{erg\,s^{-1}}$, $M_\mathrm{tot}=1.53\,M_\odot$ and $\text{T}_{eff}=4384$\,K (see Section~\ref{sec:optical_results}).

We estimate $f_\mathrm{sd} \approx 2.9$ for our system, which lies in the transition region between the ellipsoidal modulation and irradiation. The transition is expected to occur between $f_\mathrm{sd}=$2--4 \citep{2023MNRAS.525.2565T}. Figure~\ref{fig:optical_comparison} shows $L_\mathrm{sd}$ versus $P_\mathrm{orb}$ for known RB pulsars (using the data from Table (1) of  \cite{2023MNRAS.525.2565T}, along with \psr{}, illustrating that our source resides in this intermediate regime.

%We estimate $f_\mathrm{sd} \approx 2.9$ for our system, a value consistent with ellipsoidal modulation, as low $f_\mathrm{sd}$ produces light curves dominated by the companion’s shape without clear signs of irradiation. Figure~\ref{fig:optical_comparison} shows $L_\mathrm{sd}$ versus $P_\mathrm{orb}$ for known RB pulsars (using the data from Table (1) of  \cite{2023MNRAS.525.2565T}), along with \psr{}, to demonstrate how our source sits in the parameter space consistent with ellipsoidal modulation. The transition region between ellipsoidal and irradiation occurs between $f_\mathrm{sd}=2-4$. For reference, we plot $L_\mathrm{sd} \propto P_\mathrm{orb}^{4/3}$ as an approximate upper limit for hotter and more massive RB companions ($T_{eff}$= 5500\,K, $M_{tot} = 2 M_\odot$).

%In comparison, all known BWs systems have $f_{sd} > 18$, reflecting their irradiation-dominated light curves.

%This further confirms the unreliability of the LCO photometry, as in the case of ellipsoidal modulation, the variability should be observed with approximately the same amplitude across all filters.

\subsection{\textit{Fermi} Association}\label{sec:disc_fermi_prop}

\begin{figure*}
    \centering
    \includegraphics[width=\linewidth]{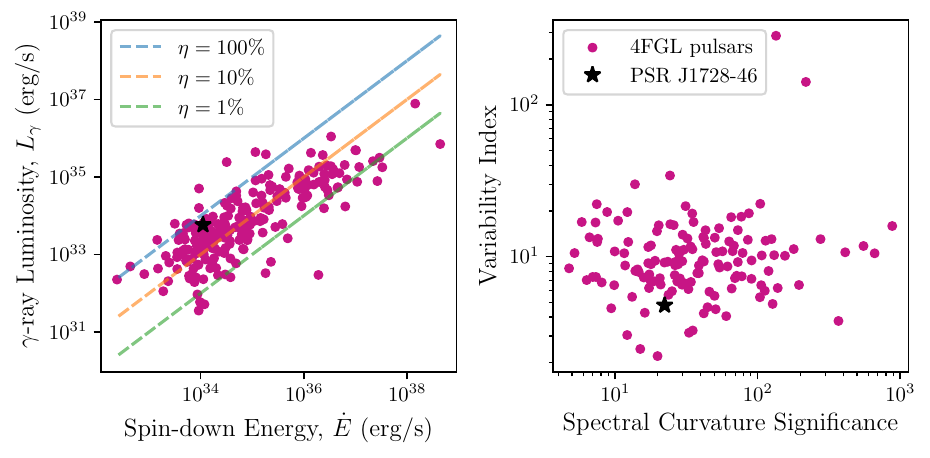}
    \caption{\textit{Left:} $L_\gamma$ vs. $\dot{E}$ for known \textit{Fermi} pulsars in the 4FGL catalogue, along with \psr. Dashed vertical lines indicate lines of constant $\gamma$-ray efficiency, $\eta$. 
    \textit{Right:} $\gamma$-ray variability index vs. spectral curvature significance for 4FGL pulsars, with \psr\ highlighted.
    See Section~\ref{sec:disc_fermi_prop} for details.}
    \label{fig:gamma_prop}
\end{figure*}

In Section~\ref{sec:Fermi} we presented the closest \textit{Fermi} $\gamma$-ray source as 4FGL\,J1728.0-4606, which lies 2$'$ from \psr{}. We analysed the properties of 4FGL\,J1728.0$-$4606 to see if they are consistent with other known $\gamma$-ray pulsars in the 4FGL catalogue. The left panel in Figure~\ref{fig:gamma_prop} shows $\dot E$ of known $\gamma$-ray pulsars (obtained by cross-matching the 4FGL catalogue with the ATNF catalogue) against $\gamma$-ray luminosity ($L_\gamma$) (from the 4FGL catalogue). $L_\gamma$ is calculated using $L_\gamma = 4\pi d^2G_{100}$, where $G_{100}$ is the measured energy flux (from 100\,MeV to\,100 GeV) available in the 4FGL catalogue, and $d$ is the distance derived from the DM of the sources in the \textsc{ATNF} catalogue. 
Using the $d$ estimated from the YMW16 model (as listed in Table~\ref{tab:timing_results}) and $G_{100}$ $= (1.01 \pm 0.10) \times 10^{-11}$\,erg\,cm$^{-2}$\,s$^{-1}$ (between 100\,MeV--100\,GeV), $L_\gamma$ of 4FGL\,J1728.0-4606 is calculated to be $5.85\times10^{33}\mathrm{erg/s}$. The figure illustrates that 4FGL\,J1728.0-4606 sits in the parameter space of other known $\gamma$-ray pulsars. We calculated a $\gamma$-ray conversion efficiency $\eta_\gamma =L_\gamma /\dot E=44\% $.

The right panel of Figure~\ref{fig:gamma_prop} shows the $\gamma$-ray spectral curvature significance against the $\gamma$-ray variability index. Pulsars are steady $\gamma$-ray emitters and are therefore expected to exhibit low variability indices (typically $<27$; \citealt{2020ApJS..247...33A}). The spectral curvature significance quantifies the deviation of the $\gamma$-ray spectrum from a simple power law and is typically greater than $3\sigma$ for pulsars, which are known to have curved spectra. Our system is consistent with the properties of previously identified RB pulsars.

\subsection{Probing The Eclipse Mechanism}\label{sec:eclipse_mec}

In the following section, we explore possible eclipse mechanisms proposed by \cite{1994ApJ...422..304T} that could be responsible for the observed eclipses in \psr{}. The size of the companions' Roche lobe($R_L$) is estimated using Equation (1) in \cite{1983ApJ...268..368E} and is found to be 0.04$R_\odot$. The eclipse radius, given by $R_E=\pi a\Delta \phi$ where $\Delta \phi = \phi_{e} - \phi_{i}$, is found to be 0.12$R_\odot$ ($\sim2.5R_L$). This suggests that the material causing the eclipses extends beyond the companion’s $R_L$.

The first eclipse mechanism we consider by \cite{1994ApJ...422..304T} is due to the plasma frequency. Eclipses can be caused when the radio wave frequency is lower than the plasma frequency in the medium. We calculate the electron density $n_e = \frac{N_e}{2R_E} = 3.4\times10^8$\,$\mathrm{cm}^{-3}$. The plasma cut-off frequency is given by $f_p = 8.5\ [\frac{n_e}{\mathrm{cm}^{-2}}]^\frac{1}{2}$\,kHz and is calculated to be 160\,MHz. Eclipses are observed above this frequency, so we rule out plasma frequency as a potential eclipse mechanism.

Eclipses due to refraction can also be ruled out, as $\sim10-100$\,ms delays are expected for refraction. 

Scattering of radio waves can cause eclipses if the pulse broadens beyond the pulsar’s spin period. We consider a total DM of 67.4856\,$\mathrm{pc\ cm^{-3}}$, consisting of the non-eclipse phase DM of 65.4856\,$\mathrm{pc\ cm^{-3}}$ plus an excess of 2.0\,$\mathrm{pc\ cm^{-3}}$ from the eclipse region. We calculate the scattering timescale to be 9.9\,$\mu$s at 1\,GHz, using Equation (7) from \cite{2015MNRAS.449.1570L}. This is substantially shorter than the pulsar’s spin period, so scattering cannot account for the eclipse, and we therefore rule out scattering as the dominant mechanism.

%In Section~\ref{sec:pulse_profile}, we calculate scattering timescales at 2368\,MHz of 0.003\,ms (using the thin screen model; \citet{1968Natur.218..920S}) and 0.00008\,ms (using the DM-based model; \citet{2004ApJ...605..759B}). Both values are significantly shorter than the pulsar’s spin period, allowing us to rule out scattering-induced eclipses.

We next consider free-free absorption as a possible eclipse mechanism. At 2368\,MHz we find $N_e = 5.9\times10^{18}$\,$\mathrm{cm}^{-2}$, and the absorption length $L=2R_E$. Using Equation (11) of \cite{1994ApJ...422..304T} and the condition that $\tau_\mathrm{ff}> 1$, we find the following condition $T\geq10^5 f_{cl}^{\frac{3}{2}}$, where $T$ is the temperature and $f_{cl}$ ($= \frac{\langle n_{e}^2 \rangle}{\langle n_{e} \rangle ^2} $) is the clumping factor of the eclipsing medium. For free-free absorption to be a valid eclipse mechanism, very low temperatures or a very high clumping factor are required. Neither is possible in this system, as the temperature of the stellar wind is typically $10^8 - 10^9$\,K, and a high $f_{cl}$ is not possible in the eclipse environment \citep{1994ApJ...422..304T}. Therefore, we rule out free-free absorption as a valid eclipse mechanism.

We consider induced Compton scattering as a possible mechanism and calculate the optical depth ($\tau_\mathrm{ICS}$) using Equation (11) of \cite{1994ApJ...422..304T}. Using the radiometer equation, we calculate a pulsed flux density of 57\,$\mu$Jy at 2368\,MHz. Using the distance of 2.2\,kpc (see Table~\ref{tab:timing_results}), a spectral index of $-1.81$ and a magnification factor $M=1$, we derive a value of $\tau _\mathrm{ind} = 0.005$. Since this value is $<<$1, induced Compton scattering can be excluded as a significant contributor to the observed eclipses.

Next, we consider cyclotron absorption. We calculate the characteristic magnetic field ($B_E$) by equating the plasma energy density $P_{B}=\frac{B_{E}^2}{8\pi}$ to the pulsar wind energy density $U_E = \frac{\dot{E}}{4\pi c a^2}$, obtaining a value of $B_E=$ 83\,G. The cyclotron frequency, $\nu_{B}=\frac{eB}{2\pi m_{e}c}$ is calculated to be approximately 230\,MHz, where $m_e$ is the mass of the electron, $e$ is the charge on the electron, and $c$ is the speed of light. The cyclotron harmonic ($m$) is found to be 10 at observing frequency 2368\,MHz, calculated using $m=\frac{\nu}{\nu_{B}}$. Using Equation (43) in \cite{1994ApJ...422..304T}, we find that $T\geq6\times10^7$\,K.  For cyclotron absorption to be valid, $T\leq \frac{m_e c^2}{2k_Bm^3} = 3\times10^6$\,K (Equation (C2) in \cite{1994ApJ...422..304T}). Therefore, we rule out cyclotron absorption as an eclipse mechanism.

Lastly, we consider synchrotron absorption, which arises from a population of non-thermal electrons, typically assumed to follow a power-law energy distribution of the form $n(E)=n_{0}E^{-p}$, where $p$ is the power-law index \citep{1994ApJ...422..304T}. The corresponding optical depth is given by \citep{1994ApJ...422..304T}:
\begin{equation}
    \tau_{\text{sync}(\nu)} = \left(\frac{ 3^\frac{(p+1)}{2} \Gamma \left( \frac{3p + 2}{12} \right) \, \Gamma \left( \frac{3p + 22}{12} \right)}{4}\right) \left( \frac{\sin \theta}{m} \right)^{\frac{p+2}{2}} \frac{n_0 e^2}{m_e c \, \nu } L
    \label{eq:sync}
\end{equation}
Where $\theta$ is the angle between the magnetic field and the line of sight, and $n_0$ is the number density of non-thermal electrons, typically taken to be $\sim$1\% of the total electron density \citep{1994ApJ...422..304T}. 

To evaluate synchrotron absorption as a potential eclipse mechanism in our system, we compute the optical depth $\tau_{\text{sync}(\nu)}$ using Equation~\ref{eq:sync} at a frequency of 2368\,MHz. We vary $p$ over the range 2--8 and $\theta$ between 0.5 and 1.5\,radians. We find that for $\tau_{\text{sync}(2368\,\mathrm{MHz})} > 1$, the parameters must satisfy $\theta > 0.61$\,radians and $p \geq 2 $. For synchrotron absorption by non-thermal electrons, $p$ can vary between 2-7, $m$ between 10--100, and $\theta$ between 0.3--1.40\,radians \citep{1982ApJ...259..350D}. For mildly relativistic electrons, $p$ is expected to fall within the range 2---3. \citep{1970SoPh...11..434T, 1969ApJ...158..753R}. In our system, the calculated values of $p$, $\theta$, and $m$ all fall within the theoretical ranges. Thus, we conclude that synchrotron absorption is the most plausible eclipse mechanism for \psr{}.

\section{Conclusions} \label{sec:conclution}

In this paper, we report the study of \psr, an RB eclipsing MSP with a spin period of 2.86 ms, a DM of 65.5\,cm$^{-3}$pc, and a 5.05\,hr orbital period with a minimum companion mass of 0.13$M_\odot$.

Timing studies revealed an $DM_{excess}$ of $2.0 \pm 1.2 \ \mathrm{pc\ cm^{-3}}$, corresponding to an electron column density of $5.9 \pm 3.6 \times10^{18} \ \mathrm{cm^{-2}}$. We find that synchrotron absorption is the most plausible eclipse mechanism in this system. Several other spider pulsars also show evidence for synchrotron absorption as the dominant process, such as PSR\,J1431$-$4715 \citep[][]{2024ApJ...973...19K} and PSR\,J1908$+$2105 \citep[][]{2025ApJ...982..168G}. Polarisation studies near the eclipse boundaries can help reveal the magnetic properties of the eclipsing medium.

\psr\ also exhibits orbital period variability, with a significant first-order orbital period derivative consistent with other known RB pulsars. This variability is likely driven by changes in the companion star’s gravitational quadrupole moment due to shape deformations over time, providing insight into the long-term evolution of the binary system.

Archival data identified a \textit{Gaia} optical counterpart within $0.5''$ of the radio position. The G-band light curve shows behaviour consistent with both ellipsoidal modulation and irradiation of the companion star. High-sensitivity optical follow-up with a $>$1-m telescope, sufficient to resolve the nearby sources, will enable study of colour evolution and further characterise the system, as indicated by the LCO data.

Finally, the detection of additional RB systems through imaging surveys, such as VAST, will expand the known sample, thereby improving our understanding of orbital dynamics, eclipsing mechanisms, and the evolutionary processes shaping these binaries.

\section{Acknowledgements}
%% Funding 
YW acknowledges the support of the Australian Research Council (ARC) grant DP220102305. 
YW acknowledges support through ARC Future Fellowship FT190100155.
 N.H.-W. is the recipient of an Australian Research Council Future Fellowship (project number FT190100231). 

%% ASKAP
This scientific work uses data obtained from Inyarrimanha Ilgari Bundara / the CSIRO Murchison Radio-astronomy Observatory. We acknowledge the Wajarri Yamaji People as the Traditional Owners and native title holders of the Observatory site. CSIRO's ASKAP radio telescope is part of the Australia Telescope National Facility (\url{https://ror.org/05qajvd42}). Operation of ASKAP is funded by the Australian Government with support from the National Collaborative Research Infrastructure Strategy. ASKAP uses the resources of the Pawsey Supercomputing Research Centre. Establishment of ASKAP, Inyarrimanha Ilgari Bundara, the CSIRO Murchison Radio-astronomy Observatory and the Pawsey Supercomputing Research Centre are initiatives of the Australian Government, with support from the Government of Western Australia and the Science and Industry Endowment Fund.
%% ATCA
The Australia Telescope Compact Array is part of the Australia Telescope National Facility (\url{https://ror.org/05qajvd42}), which is funded by the Australian Government for operation as a National Facility managed by CSIRO.
We acknowledge the Gomeroi people as the Traditional Owners of the Observatory site.
%% Parkes
Murriyang, the Parkes radio telescope, is part of the Australia Telescope National Facility (\url{https://ror.org/05qajvd42}), which is funded by the Australian Government for operation as a National Facility managed by CSIRO.
We acknowledge the Wiradjuri people as the Traditional Owners of the Observatory site.
%% OzGrav
Parts of this research were conducted by the Australian Research Council Centre of Excellence for Gravitational Wave Discovery (OzGrav), through project number CE230100016. 
%% Shanghai HPC
This work used resources of China SKA Regional Centre prototype \citep{An2019NatAs...3.1030A,An2022SCPMA..6529501A} funded by National SKA Program of China (2022SKA0130103) and the National Key R\&D Programme of China (2018YFA0404603). 
%% Jinan HPC 
%% VizieR
This research has made use of the VizieR catalogue access tool, CDS, Strasbourg, France (DOI: \url{10.26093/cds/vizier}). 
The original description of the VizieR service was published in \citet{2000A&AS..143...23O}.

% MWA
%This scientific work uses data obtained from Inyarrimanha Ilgari Bundara, the CSIRO Murchison Radio-astronomy Observatory. Support for the operation of the MWA is provided by the Australian Government (NCRIS), under a contract to Curtin University administered by Astronomy Australia Limited. ASVO has received funding from the Australian Commonwealth Government through the National eResearch Collaboration Tools and Resources (NeCTAR) Project, the Australian National Data Service (ANDS), and the National Collaborative Research Infrastructure Strategy.
% Pawsey
We acknowledge the Pawsey Supercomputing Centre, which is supported by the Western Australian and Australian Governments.
% MeerKAT
The MeerKAT telescope is operated by the South African Radio Astronomy Observatory, which is a facility of the National Research Foundation, an agency of the Department of Science and Innovation.
This work has made use of the ``MPIfR S-band receiver system'' designed, constructed and maintained with funding of the MPI für Radioastronomy and the Max-Planck-Society. Observations made use of the Pulsar Timing User Supplied Equipment (PTUSE) servers at MeerKAT, which were funded by the MeerTime Collaboration members ASTRON, AUT, CSIRO, ICRAR-Curtin, MPIfR, INAF, NRAO, Swinburne University of Technology, the University of Oxford, UBC and the University of Manchester. The system design and integration was led by Swinburne University of Technology and Auckland University of Technology in collaboration with SARAO and supported by the ARC Centre of Excellence for Gravitational Wave Discovery (OzGrav) under grant CE170100004.
% uGMRT
We thank the staff of the GMRT who made these observations possible. GMRT is run by the National Centre for Radio Astrophysics of the Tata Institute of Fundamental Research.

\vspace{5mm}
%\facilities{ASKAP, ATCA, Parkes, MeerKAT, uGMRT}


%\software{
%aplpy~\citep{Robitaille2012ascl.soft08017R}, 
%astropy~\citep{AstropyCollaboration2013A&A...558A..33A,AstropyCollaboration2018AJ....156..123A}, 
%astroquery~\citep{Ginsburg2019AJ....157...98G}, 
%matplotlib~\citep{Hunter2007CSE.....9...90H},
%numpy~\citep{Harris2020Natur.585..357H}, 
% \textsc{pygsm}~\citep{Price2016}, 
%and scipy~\citep{Virtanen2020NatMe..17..261V}. 
%}

%\bibliography{ref,cite}{}
%\bibliographystyle{aasjournal}
%\printbibliography

\bibliography{J1728}
\end{document}